\documentclass[journal,transmag]{IEEEtran}

\usepackage[english]{babel}
\usepackage{amsmath,amsthm}
\usepackage{amsfonts}
\usepackage{xspace}
\usepackage{bm}
\usepackage[squaren, thinqspace]{SIunits}
\usepackage{booktabs}
\usepackage{graphicx}

\theoremstyle{definition}

\theoremstyle{remark}

\usepackage{cite}
\begin{document}
\renewcommand{\degree}{\ensuremath{^\circ}\xspace}
\renewcommand{\Re}{\ensuremath{\mathrm{Re}} \xspace}
\renewcommand{\Im}{\ensuremath{\mathrm{Im}} \xspace}
\newcommand{\Hy}{\ensuremath{\bm{H} || \bm{y}}\xspace}
\newcommand{\Hx}{\ensuremath{\bm{H} || \bm{x}}\xspace}
\newcommand{\y}{\ensuremath{\bm{y}}\xspace}
\newcommand{\x}{\ensuremath{\bm{x}}\xspace}
\newcommand{\z}{\ensuremath{\bm{z}}\xspace}
\newcommand{\ii}{\ensuremath{\mathrm{i}}\xspace}
\newcommand{\Vdc}{\ensuremath{V_\mathrm{dc}}\xspace}
\newcommand{\Vdcs}{\ensuremath{V_\mathrm{dc}^\mathrm{s}}\xspace}
\newcommand{\Vdca}{\ensuremath{V_\mathrm{dc}^\mathrm{a}}\xspace}
\newcommand{\Vsp}{\ensuremath{V_\mathrm{SP}}\xspace}
\newcommand{\Vvna}{\ensuremath{V_1}\xspace}
\newcommand{\gSpinMix}{\ensuremath{g_{\uparrow\!\downarrow}}\xspace}
\newcommand{\VISHEdc}{\ensuremath{V_\mathrm{dc}^\mathrm{iSHE}}\xspace}
\newcommand{\VISHEac}{\ensuremath{V_\mathrm{iSHE}}\xspace}
\newcommand{\Js}{\ensuremath{\bm{J}_\mathrm{s}}\xspace}
\newcommand{\mus}{\ensuremath{\bm{\mu}_\mathrm{s}}\xspace}
\newcommand{\Jc}{\ensuremath{\bm{J}_\mathrm{c}}\xspace}
\newcommand{\alphaSH}{\ensuremath{\Theta_{\mathrm{SH}}}\xspace}
\newcommand{\lambdaSD}{\ensuremath{\lambda_{\mathrm{SD}}}\xspace}
\newcommand{\tC}{\ensuremath{t_{\mathrm{Cu}}}\xspace}
\newcommand{\sigmaC}{\ensuremath{\sigma_{\mathrm{Cu}}}\xspace}
\newcommand{\tN}{\ensuremath{t_{\mathrm{N}}}\xspace}
\newcommand{\sigmaN}{\ensuremath{\sigma_{\mathrm{N}}}\xspace}
\newcommand{\tF}{\ensuremath{t_{\mathrm{F}}}\xspace}
\newcommand{\tPy}{\ensuremath{t_{\mathrm{Py}}}\xspace}
\newcommand{\sigmaF}{\ensuremath{\sigma_{\mathrm{F}}}\xspace}
\newcommand{\SP}{\ensuremath{\bm{\hat{s}}}\xspace}
\newcommand{\hrf}{\ensuremath{\bm{h}_\mathrm{mw}}\xspace}
\newcommand{\hmw}{\ensuremath{h_\mathrm{mw}}\xspace}
\newcommand{\Jrf}{\ensuremath{\bm{I}_\mathrm{mw}}\xspace}
\newcommand{\HexB}{\ensuremath{\bm{H}_0}\xspace}
\newcommand{\Hex}{\ensuremath{H_0}\xspace}
\newcommand{\Heff}{\ensuremath{\bm{H}_\mathrm{eff}}\xspace}
\newcommand{\M}{\ensuremath{\bm{M}}\xspace}
\newcommand{\my}{\ensuremath{m_y}\xspace}
\newcommand{\hy}{\ensuremath{h_y}\xspace}
\newcommand{\mz}{\ensuremath{m_z}\xspace}
\newcommand{\My}{\ensuremath{M_y}\xspace}
\newcommand{\Mz}{\ensuremath{M_z}\xspace}
\newcommand{\Ms}{\ensuremath{M_\mathrm{s}}\xspace}
\newcommand{\Vind}{\ensuremath{V_\mathrm{FMI}}\xspace}
\newcommand{\Eind}{\ensuremath{\bm{E}_\mathrm{ind}}\xspace}
\newcommand{\Eishe}{\ensuremath{\bm{E}_\mathrm{iSHE}}\xspace}
\newcommand{\muBohr}{\ensuremath{\mu_\mathrm{B}}\xspace}
\newcommand{\Hres}{\ensuremath{H_\mathrm{res}}\xspace}
\newcommand{\Ztot}{\ensuremath{A}\xspace}
\newcommand{\Zishe}{\ensuremath{A_\mathrm{iSHE}}\xspace}
\newcommand{\Zind}{\ensuremath{A_\mathrm{FMI}}\xspace}
\newcommand{\Zdc}{\ensuremath{V_0^\mathrm{iSHE}}\xspace}
\newcommand{\Phifit}{\ensuremath{\phi_\mathrm{fit}}\xspace}
\newcommand{\Phiref}{\ensuremath{\phi_\mathrm{fit}^\mathrm{ref}}\xspace}
\newcommand{\Phiishe}{\ensuremath{\phi_\mathrm{iSHE}}\xspace}
\newcommand{\Phiind}{\ensuremath{\phi_\mathrm{FMI}}\xspace}
\newcommand{\Phitot}{\ensuremath{\phi}\xspace}
\newcommand{\Phiindy}{\ensuremath{\phi_\mathrm{FMI}^{y}}\xspace}
\newcommand{\chiT}{\ensuremath{\bm{\chi}}\xspace}
\newcommand{\Pmw}{\ensuremath{P_\mathrm{mw}}\xspace}

%
\title{Detection of the dc inverse spin Hall effect due to spin pumping in a novel meander-stripline geometry}


\author{\IEEEauthorblockN{Mathias Weiler\IEEEauthorrefmark{},
Justin M. Shaw\IEEEauthorrefmark{},
Hans T. Nembach\IEEEauthorrefmark{}, and
Thomas J. Silva\IEEEauthorrefmark{}}%
\IEEEauthorblockA{\IEEEauthorrefmark{}Electromagnetics Division, National Institute of Standards and Technology, Boulder, CO, 80305 \\
Contribution of NIST, not subject to copyright}
\thanks{Corresponding author: Mathias Weiler (mathias.weiler@nist.gov)}}

\markboth{Microwave Magnetics}{Microwave Magnetics}

\IEEEtitleabstractindextext{%
\begin{abstract}
The dc voltage obtained from the inverse spin Hall effect (iSHE) due to spin pumping in ferromagnet/normal-metal (NM) bilayers can be unintentionally superimposed with magnetoresistive rectification of ac charge currents in the ferromagnetic layer. We introduce a geometry in which these spurious rectification voltages vanish while the iSHE voltage is maximized. In this geometry, a quantitative study of the dc iSHE is performed in a broad frequency range for Ni$_{80}$Fe$_{20}$/NM multilayers with NM=\{Pt, Ta, Cu/Au, Cu/Pt\}. The experimentally recorded voltages can be fully ascribed to the iSHE due to spin pumping. Furthermore we measure a small iSHE voltage in single CoFe thin films.
\end{abstract}

\begin{IEEEkeywords}
Microwave magnetics, spin pumping, spin rectification, inverse spin Hall effect
\end{IEEEkeywords}}

\maketitle

\IEEEdisplaynontitleabstractindextext

\IEEEpeerreviewmaketitle

\section{Introduction}
Spin pumping is a significant source of damping in ultra-thin ferromagnet/normal-metal (FM/NM) bilayers \cite{Tserkovnyak:2002}. In these bilayers, the precessing magnetization of the ferromagnet relaxes partially via the emission of ac transverse and dc longitudinal spin currents. Thus, magnetization dynamics in FM/NM junctions are a source of pure spin currents for spintronic applications~\cite{Zutic:2004, Bader:2010}. A large number of experimental and theoretical works are concerned with achieving a quantitative understanding of the spin pumping process~\cite{Azevedo:2005,Costache:2006,Mosendz:2010a, Mosendz:2010, Czeschka:2011,Jiao:2013}. The spin currents can be detected electrically via the inverse spin Hall effect  (iSHE)~\cite{Dyakonov:1971, Hirsch:1999,Saitoh:2006} in the normal metal. However, electrical spin-current detection is complicated by the presence of inductive signals in the ac regime~\cite{Weiler:2014a} and voltages due to microwave rectification in the dc regime~\cite{Juretschke:1960,Egan:1963,Gui:2007,Mecking:2007,Yamaguchi:2007,Azevedo:2011}. Considerable efforts have been made to separate the dc spin pumping voltages from spurious rectification signals via line shape analysis~\cite{Mosendz:2010a, Mosendz:2010} or by studies where the external magnetic field orientation is changed in a prescribed way~\cite{Bai:2013}. Here, we experimentally demonstrate the use of an optimized sample geometry that entirely eliminates these voltage rectification effects. We determine the effective spin-mixing conductance from broad-band ferromagnetic resonance (FMR) measurements and extract the dc spin Hall angles for Pt, Ta and Au from the recorded dc iSHE voltages. A dc voltage in ferromagnetic resonance (FMR) is also observed in a single Co$_{90}$Fe$_{10}$ (CoFe) layer, consistent with the hypothesis of a self-induced, inverse spin Hall effect in this alloy.

\section{Sample design and preparation}
\begin{figure}
  \centering
  \includegraphics{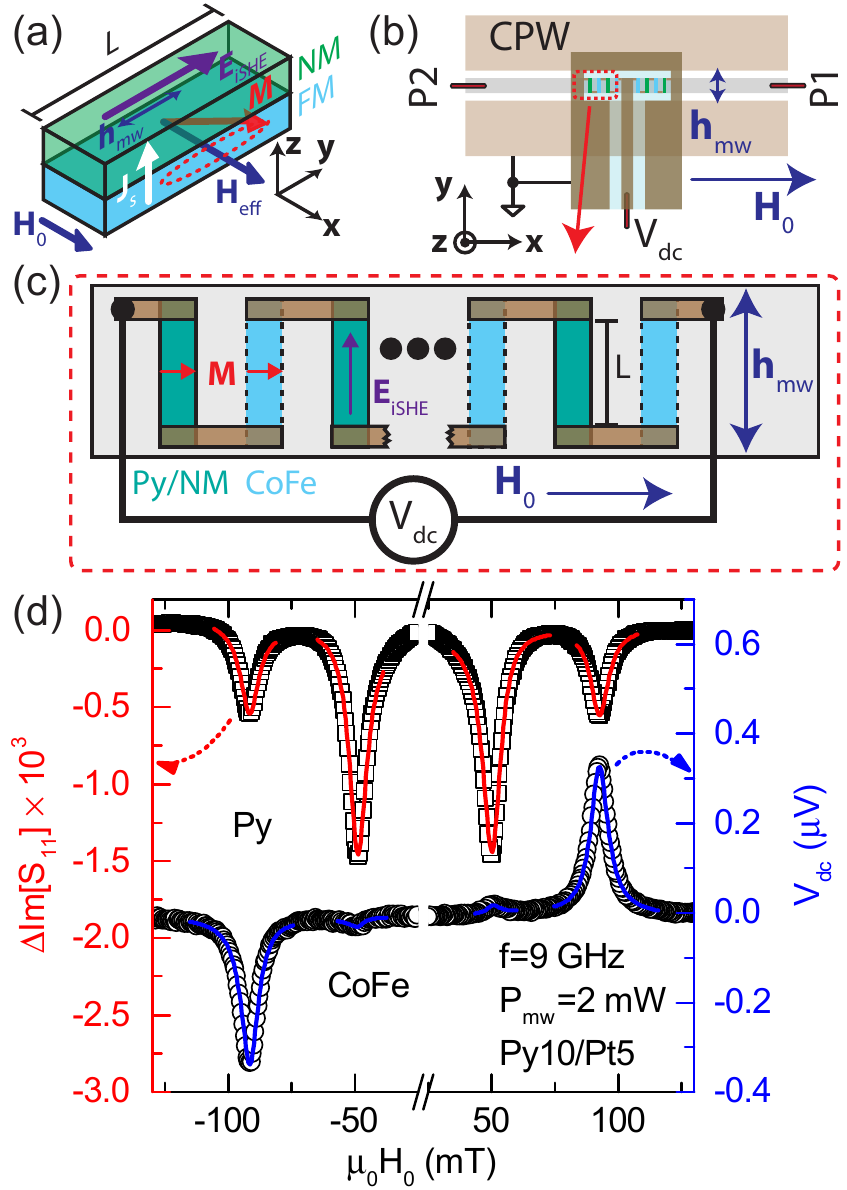}\\
  \caption{(a) Geometry for measurement of spin pumping via the dc iSHE in FM/NM bilayers. (b) Sketch of the devices used. (c) Closeup showing the meander-arrangement of the total of $N=10$ Py/NM bilayer and $N=10$ CoFe single layer stripes on top of the CPW center conductor. (d) $S_{11}$ data (left scale; only imaginary part shown)  versus external magnetic field \Hex and simultaneously acquired dc voltage \Vdc (right scale) at $f=\unit{9}{\giga\hertz}$.}\label{fig:device}
\end{figure}
The geometry for iSHE detection of the dc component of the pumped spin current is shown in Fig.~\ref{fig:device}(a). We use a FM/NM bilayer with the interface normal along \z, where the magnetic layer is a polycrystalline film with easy-plane anisotropy. The external magnetic field $\HexB\parallel\x$ is applied orthogonal to the ac driving field $\hrf\parallel\y$. This allows for the most efficient excitation of the magnetization \M with its equilibrium orientation along $\Heff\parallel\x$. The dc component of the spin current \Js has spin polarization \SP along $\Heff$. The open-circuit electric field $\Eishe\propto\Js\times\SP$ due to the inverse spin Hall effect in the NM is oriented along \y. The corresponding dc iSHE voltage is $\VISHEdc=E_\mathrm{iSHE}  L $, where $L$ is the length of the bilayer along \y. 

In our broadband experiments, \hrf represents an Oersted field generated by a microwave charge current $\Jrf\parallel\x$ applied to the $w_\mathrm{CPW}=\unit{150}{\micro\meter}$-wide center conductor of the coplanar waveguide (CPW), shown in Fig.~\ref{fig:device}(b). Importantly, the spurious resonant rectification voltage due to anisotropic magnetoresistance that occurred in previous studies~\cite{Mosendz:2010a, Mosendz:2010} vanishes in this geometry because \HexB is applied along a highly symmetric axis within a precision of $\pm 2\degree$. To take advantage of this measurement geometry, all samples are patterned into a meander-line structure, sketched in Fig.~\ref{fig:device}(c). Every other strip consists of identical thin-film Py/NM bilayers (Py=Ni$_{80}$Fe$_{20}$), while the remaining strips are single-layer \unit{15}{\nano\meter}-thick CoFe films. All bilayers are $w=\unit{25}{\micro\meter}$ wide and $L=\unit{100}{\micro\meter}$ long. Contacts between the individual bilayers consist of \unit{180}{\nano\meter}-thick Cu films capped with \unit{20}{\nano\meter} Au. The iSHE voltages in all Py/NM and CoFe strips add and the measured total dc voltage \Vdc scales linearly with the number $N$ of meander repeats. For all data shown here, $N=10$. All thin films are prepared by sputter deposition without breaking the vacuum between deposition of Py and NM. The samples are placed with the meander-line on top of the CPW center conductor with an air gap of $\delta\approx\unit{50}{\micro\meter}$ and the NM facing the CPW.

\section{Results and discussion}
We use a 2-port vector network analyzer (VNA) with output power $\Pmw=\unit{2}{\milli\watt}$ that is connected and calibrated to ports P1 and P2 of the device sketched in Fig.~\ref{fig:device}(b). The VNA excites the CPW and inductively detects the magnetization dynamics. Figure~\ref{fig:device}(d) shows raw data obtained with a meander line composed of Py10/Pt5 bilayers and CoFe single layers. (Integer numbers in the sample names are nominal layer thicknesses in nanometers). The open squares (left scale) are  the imaginary part of $S_{11}$  acquired with the VNA at a fixed frequency $f=\unit{9}{\giga\hertz}$ as a function of \Hex. The $S_{11}$ data shows two distinct resonances, one at $\mu_0|\Hex|\approx\unit{40}{\milli\tesla}$ and one at $\mu_0|\Hex|\approx\unit{80}{\milli\tesla}$. The resonances in $S_{11}$ are attributed to the inductive detection of the FMR of the CoFe thin films and Py/NM bilayers, respectively. Thus, the field-swept $S_{11}$ data show a change $\Delta S_{11}=\Vind/V_0$ at the resonance fields of both the CoFe and the Py/NM bilayers, where the inductive voltage \Vind is defined in~\cite{Weiler:2014a} and $V_0$ is the ac voltage applied to the CPW. The higher saturation magnetization \Ms of CoFe, compared to that of Py, results in the smaller absolute in-plane resonance field for CoFe. The CoFe resonance is more pronounced because the CoFe layer is thicker (\unit{15}{\nano\meter} CoFe vs.~\unit{10}{\nano\meter} Py) and \Ms is higher. The simultaneously-recorded dc voltage \Vdc (circles, right scale) exhibits a large resonance at the same field, where the inductive signal has a dip due to the Py response. The resonant value of \Vdc inverts polarity with \Hex inversion, as expected for the detection of a dc spin current \Js via the iSHE. Unexpectedly, small \Vdc resonance features also appear for CoFe. We present further analysis of these features below. First, we discuss the $S_{11}$ and \Vdc signals at the Py resonance.
\begin{figure}
  \centering
  \includegraphics{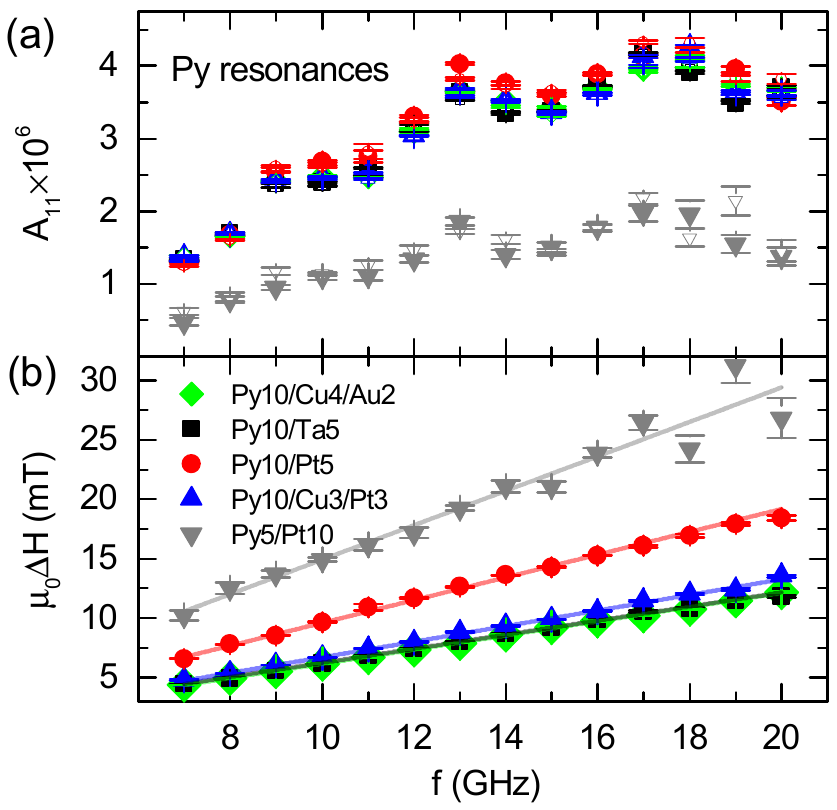}\\
  \caption{(color online) (a) The resonance magnitude $A_{11}$ of the inductive signal obtained from fits to the $S_{11}$ data is similar for all samples with \unit{10}{\nano\meter}-thick Py. $A_{11}$ obtained from the Py5/Pt10 sample is a factor 2 smaller. (b) Linear fits to the field-swept linewidths of all Py/NM resonances are used to extract $\Delta H_0$ and $\alpha$.}\label{fig:Chi}
\end{figure}
\begin{table*}
\renewcommand{\arraystretch}{1.3}
\centering
\caption{Fitted parameters for the Py/NM samples used in this study.}
\label{tab:samples}
\begin{tabular}{lccccccc}
  Sample & $M_\mathrm{eff}$ (kA/m) & $ g$ & $\mu_0 \Delta H_0$ (mT) & $\alpha$ & \gSpinMix ($10^{19}$ m$^{-2}$)  & \alphaSH\\ 
\hline   
Py10/Cu4/Au2 & $732\pm1$ & $2.106\pm0.001$ & $-0.2\pm0.1$ & $0.0088\pm0.0001$ & $1.44\pm0.05$  &$0.005\pm0.001$\\%
Py10/Ta5 & $753\pm2$ & $2.113\pm0.002$ & $0.3\pm0.09$ & $0.0088\pm0.0001$ & $1.44\pm0.05$ & $-0.018\pm0.001$\\%
Py10/Pt5 & $731\pm5$ & $2.113\pm0.005$ & $0.0\pm0.1$ & $0.0142\pm0.0002$ & $4.21\pm0.10$  &$0.107\pm0.003$\\%
Py10/Cu3/Pt3 & $740\pm2$ & $2.106\pm0.002$ & $0.2\pm0.1$ & $0.0097\pm0.0001$ & $1.90\pm0.05$  &$0.116\pm0.004$\\
Py5/Pt10 & $626\pm3$ & $2.134\pm0.03$ & $-1.0\pm0.9$ & $0.0232\pm0.001$ & $4.37\pm0.03$  &$0.096\pm0.003$\\
\end{tabular}
\end{table*}

We fit the complex $S_{11}$ data at the Py and CoFe resonances with the Polder susceptibility $\chi_{yy}$~\cite{Nembach:2011,Weiler:2014a} and perform Levenberg-Marquardt optimization of $S_{11}(H_0)=C_1+\Hex C_2+A_{11}\cdot\chi_{yy}(H_0,\Hres,\Delta H,\phi)$. The resultant fits are shown by the lines in the $S_{11}$ data of Fig.~\ref{fig:device}(d). Parameters of the fits are the resonance field \Hres, the linewidth $\Delta H$, the magnitude $A_{11}$, the phase $\phi$, and the complex field-independent offset and slope $C_1$ and $C_2$, respectively. $A_{11}$ is shown in Fig.~\ref{fig:Chi}(a) for all Py/NM multilayers investigated in this study. Solid symbols are for $\Hex>0$ and open symbols for $\Hex<0$. $A_{11}$ is very similar for both \Hex directions and all samples except for Py5/Pt10, where $A_{11}$ is reduced by a factor of 2 compared with the samples of \unit{10}{\nano\meter}-thick Py, in agreement with the inductive detection mechanism. By virtue of reciprocity, the $A_{11}$ data in Fig.~\ref{fig:Chi}(a) suggest identical values of \hrf for all samples.
From a Kittel fit to \Hres~\cite{Weiler:2014a} for all investigated Py/NM bilayers, we extract both the effective magnetization $M_\mathrm{eff}=\Ms-H_\mathrm{k}^\perp$, and the Land\'{e} factor $g$ for Py. We then determine the Gilbert damping constant $\alpha$ and the inhomogeneous broadening $\Delta H_0$ from the slope and intercept, respectively, of the linear fits to the linewidth $\Delta H$ vs.~$f$, as shown in Fig.~\ref{fig:Chi}(b). The effective spin mixing conductance 
\begin{equation}\label{eq:gSpinMix}
\gSpinMix=\left(\alpha-\alpha_0\right)\frac{4\pi \Ms \tF}{\hbar \gamma}
\end{equation}
is extracted by use of $\alpha_0=0.006$ and $\Ms=\unit{800}{\kilo\ampere\per\meter}$ obtained from a nominally \unit{10}{\nano\meter}-thick reference Py film by FMR and SQUID magnetometry, respectively. Here, \tF is the Py thickness and $\gamma=g\muBohr/\hbar$ is the gyromagnetic ratio with the reduced Planck constant $\hbar$ and the Bohr magneton $\muBohr$. The parameters obtained from the fits are summarized in Table~\ref{tab:samples}.

\begin{figure}
  \centering
  \includegraphics{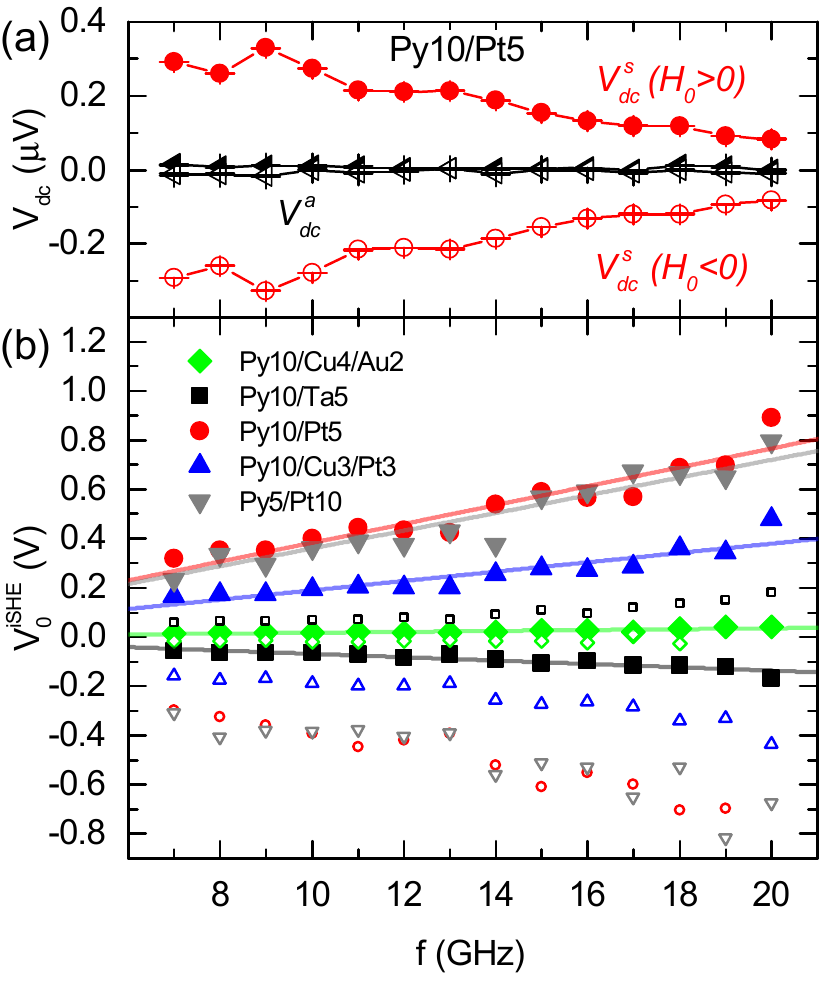}\\
  \caption{(color online) (a) The fitted symmetric (\Vdcs, circles) and antisymmetric (\Vdca, triangles) contributions to the dc voltage \Vdc recorded with the Py10/Pt5 sample show vanishing \Vdca while \Vdcs changes sign under inversion of \Hex. This indicates that \Vdc is entirely due to the detection of spin pumping via the iSHE. (b) \Zdc (closed symbols $\Hex>0$, open symbols $\Hex<0$) is calculated for all samples according to Eq.~\eqref{eq:Zdc}. Lines are fits to Eq.~\eqref{eq:theory}.}\label{fig:Vdc}
\end{figure}
The following three conditions need to be fulfilled if \Vdc is entirely due to detection of spin-pumping via the iSHE: (a) \Vdc has a purely symmetric Lorentzian line shape, (b) \Vdc changes sign under \Hex inversion, and (c) the magnitude of \Vdc is unchanged under \Hex inversion.  We thus fit the \Vdc data to the superposition of symmetric and antisymmetric Lorentzian line shapes
\begin{equation}\label{eq:gSpinMix}
\Vdc=\frac{\Vdcs\Delta^2 +\Vdca \left[\Delta\left(\Hex-\Hres\right)\right]}{\Delta^2+\left(\Hex-\Hres\right)^2}+C_3+H_0C_4\;,
\end{equation}
where $\Delta=\Delta H/2$. $C_3$ and $C_4$ describe offset and drift in \Vdc. The extracted \Hres and $\Delta H$ coincide with those extracted from the $S_{11}$ fit. The fitted symmetric (\Vdcs) and antisymmetric (\Vdca) contributions of the Py10/Pt5 resonance are shown in Fig.~\ref{fig:Vdc}(a) where $\Vdca=0$ to within uncertainty of the fit and \Vdcs changes sign with inversion of \Hex  while  $|\Vdcs|$ remains unchanged [cf.~full and open circles in Fig.~\ref{fig:Vdc}(a)]. This is consistent with our interpretation that \Vdc is entirely due to the iSHE detection of spin pumping, as expected from the chosen experimental geometry. For a quantitative analysis, we introduce the normalized iSHE voltage
\begin{equation}\label{eq:Zdc}
\Zdc=\frac{\Vdcs \Ms^2}{|\chi_{yy}(\Hres)||\chi_{zy}(\Hres)|\hmw^2}\;
\end{equation}
with the susceptibility $\chi$ obtained from the fits to $S_{11}$. \Zdc is normalized to the magnetization precession cone angle and thus is a direct measure for the iSHE efficiency for a given Py/NM stack.  In~\eqref{eq:Zdc}, \hmw is calculated from the inductive signal magnitude $A_{11}$ in Fig.~\ref{fig:Chi}(a) by
\begin{equation}\label{eq:hmw}
\hmw=\frac{4 \sqrt{\Pmw Z_0} A_{11}}{2 \pi f \mu_0 w N \tF \eta}\;, 
\end{equation}
where $Z_0=\unit{50}{\ohm}$ and $\Pmw=\unit{2}{\milli\watt}$. Here, $\eta=2 \arctan(w_\mathrm{CPW}/(2\delta))/\pi$ accounts for the non-zero spacing between the CPW and the meander line. By inferring \hmw from the measured $A_{11}$,  variations of \hmw with frequency, due to non-idealities of the loaded CPW, are quantitatively taken into account.

From spin pumping theory, we expect~\cite{Jiao:2013}
\begin{equation}\label{eq:theory}
\Zdc=\frac{f L N e\gSpinMix\alphaSH\lambdaSD \tanh\left(\frac{\tN}{2\lambdaSD}\right)}{\left(\tF\sigmaF+\tN\sigmaN+\tC\sigmaC\right)}\;,
\end{equation}
where $e$ is the electron charge, $\tF/\tC/\tN$ are the thickness of ferromagnet/Cu/normal metal layers and $\sigmaF/\sigmaC/\sigmaN$ are the corresponding electrical conductivities. In~\eqref{eq:theory}, \alphaSH and \lambdaSD are the spin Hall angle and spin diffusion length of the normal metal, respectively. We assume that the iSHE of Cu is negligible and $\lambdaSD^\mathrm{Cu}\gg\tC$, such that the insertion of Cu in the stack will change only \gSpinMix and the multilayer resistance. 
\begin{figure}
  \centering
  \includegraphics{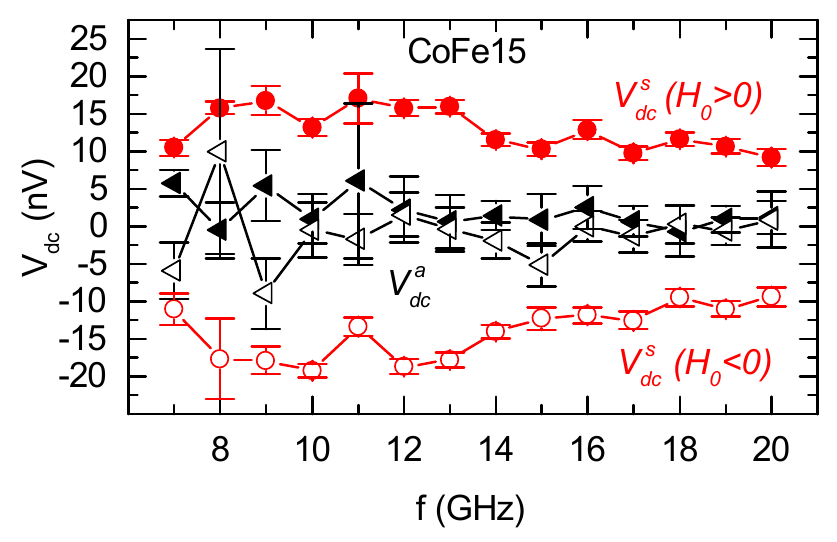}\\
  \caption{(color online) \Vdc at the CoFe resonance. Within error, $\Vdca=0$ (triangles) while \Vdcs (circles) changes sign under inversion of \Hex. \Vdcs is attributed to the detection of spin pumping via the iSHE in a single layer of CoFe due to a nonuniform dynamic magnetization.}\label{fig:CoFe}
\end{figure}
The lines in Fig.~\ref{fig:Vdc}(b) are fits to Eq.~\eqref{eq:theory} where values of electrical conductances are obtained from dc resistance measurements, and $\lambdaSD^\mathrm{Au}=\unit{34}{\nano\meter}$ from~\cite{Mosendz:2010}, $\lambdaSD^\mathrm{Pt}=\unit{1.5}{\nano\meter}$ from~\cite{Weiler:2013} and $\lambdaSD^\mathrm{Ta}=\unit{1}{\nano\meter}$ from~\cite{Boone:2013} are used. From the fits, we obtain \alphaSH in Table~\ref{tab:samples}. In particular, $\alphaSH$ for both Py/Pt samples and Py/Cu/Pt are nearly identical. Thus, we find no indication for either proximity-induced contributions to the dc voltage in Py/Pt or substantial change of interfacial spin flip for the dc spin-current by insertion of Cu. The negligibly small \Zdc in Py/Cu/Au is consistent with a weak/non-existent iSHE for both Cu and Au, as assumed in~\cite{Weiler:2014a}. 

In Fig.~\ref{fig:CoFe}, we plot the extracted \Vdca and \Vdcs for the CoFe resonance. The interspersed multilayer stripes are Py10/Pt5 for this particular sample. Identical CoFe signals were obtained for all samples of this study. Within the fit uncertainty, $\Vdca=0$. However, a small but detectable \Vdcs changes sign under inversion of \Hex as shown by the open and closed symbols in Fig.~\ref{fig:CoFe}. We attribute \Vdcs to the occurance of the dc iSHE in the single layer of CoFe itself, similar to what was recently reported for a single layer of Py~\cite{Tsukahara:2013}. Such a "self-induced" iSHE requires some degree of symmetry breaking through the film thickness. We conjecture that the CoFe layer has a nonuniform dynamic magnetization profile through the film thickness due to nonzero conductivity effects~\cite{Maksymov:2013, Weiler:2014a}. Thus, antisymmetric intralayer spin-currents are not balanced and can give rise to a net iSHE dc voltage.  
Because the nonuniform magnetization profile in CoFe could not be determined, a value for the spin Hall angle for CoFe cannot be estimated from the measured \Vdcs. However, ferromagnetic metals are known to generally display an iSHE~\cite{Miao:2013}.

\section{Summary}
We have demonstrated a sample geometry that allows unambiguous, broadband detection of spin pumping via the dc-voltage due to the iSHE. An optimized meandering structure is used to suppress rectified voltages due to anisotropic magnetoresistance. \Vdc can easily be enhanced by increasing the number of repeats of the meander structure, and the high signal-to-noise ratio allows for detection of \Vdc signals in the \unit{10}{\nano\volt} range. Among other benefits, this structure allows for the investigation of small iSHE effects at relatively low microwave powers that do not either heat the sample or induce undesired nonlinear spin dynamics. Experimental evidence supports a self-induced iSHE due to intralayer spin pumping in a single-layer \unit{15}{\nano\meter}-thick CoFe film.

\section*{Acknowledgment}
M.W. acknowledges a stipend of the German Academic Exchange Service (DAAD) and thanks Emilie Ju\'{e} for valuable comments during the manuscript preparation.

\ifCLASSOPTIONcaptionsoff
  \newpage
\fi


\end{document}